\documentclass[%
    final,
    notitlepage,
    narroweqnarray,
    inline,
    twoside,
    ]{ieee}

\begin{document}

\title[Web Server Benchmark Application]{%
       Web Server Benchmark Application \textit{WiiBench} using Erlang/OTP R11 and Fedora-Core Linux 5.0}

\author[MUTIARA AND SABASTIAN]{Achmad B. Mutiara\member{Regular
       Member},\authorinfo{A.\,B.\,Mutiara is with the Graduate Program in Information System,
       Gunadarma University, Depok 16424, Indonesia.
       E-mail: amutiara@staff.gunadarma.ac.id}%
\and{}and T.A. Sabastian \authorinfo{T.\,A.\,Sabastian is with
       the Department of Informatics Engineering, Gunadarma University, Depok 16424, Indonesia.
       } }



\maketitle

\begin{abstract}
As the web grows and the amount of traffics on the web server
increase, problems related to performance begin to appear. Some of
the problems, such as the number of users that can access the
server simultaneously, the number of requests that can be handled
by the server per second (requests per second) to bandwidth
consumption and hardware utilization like memories and CPU. To
give better quality of service (\textbf{\textit{QoS}}), web
hosting providers and also the system administrators and network
administrators who manage the server need a benchmark application
to measure the capabilities of their servers. Later, the
application intends to work under Linux/Unix -- like platforms and
built using Erlang/OTP R11 as a concurrent oriented language under
Fedora Core Linux 5.0. \textbf{\textit{WiiBench}} is divided into
two main parts, the controller section and the launcher section.
Controller is the core of the application. It has several duties,
such as read the benchmark scenario file, configure the program
based on the scenario, initialize the launcher section, gather the
benchmark results from local and remote Erlang node where the
launcher runs and write them in a log file (later the log file
will be used to generate a report page for the sysadmin).
Controller also has function as a timer which act as timing for
user inters arrival to the server. Launcher generates a number of
users based on the scenario, initialize them and start the
benchmark by sending requests to the web server. The clients also
gather the benchmark result and send them to the controller.

\end{abstract}


\section{Introduction}

\PARstart In the last two decades, human necessities in fast and
accurate information create a lot of innovations in information
technology, one of them is the internet. Since TCP/IP released to
public in 1982 and World Wide Web (WWW) introduced in 1991,
internet has became a popular media to access and publish
information. The easy to use web mechanisms make people easy to
search and publish information on the internet. The web service
later grows to many aspects, such as entertainment, education,
scientific research and many more.

To access the web on the internet, we need a certain server than
can provide user access on the web pages. This server is called
web server or HTTP server and has a main duty to serve user access
to web pages contents, either static or dynamic.

As the web grows and the amount of traffics on the web server
increase, problems related to performance begin to appear. Some of
the problems, such as the number of users that can access the
server simultaneously, the number of requests that can be handled
by the server per second (requests per second) to bandwidth
consumption and hardware utilization like memories and CPU.

To give better quality of service (QoS) , web hosting providers
and also the system administrators and network administrators who
manage the server need a benchmark application to measure the
capabilities of their servers. Later, the application intends to
work under Linux/Unix -- like platforms and built using Erlang/OTP
R11 as a concurrent oriented language under Fedora Core Linux 5.0.

Base on the above descriptions, there are some problems than can be summarized, such as:

\begin{enumerate}
\item To give better Quality of Service, web hosting provider and
also system administrators and network administrators who manage
the web server need a benchmark application to measure the
capabilities/ performances of their servers.

\item The benchmark application is intended to be use by the
network administrators and system administrators who work under
Linux/Unix -- like systems.

\item The application is made by utilizing the concurrent
capability of Erlang programming language under Linux operating
system.
\end{enumerate}

\section{Theories}

\subsection{Web Server}

The term web server can mean one of two things [2]:

\begin{enumerate}
\item A computer or a number of computers which responsible for
accepting HTTP requests from clients, which are known as web
browsers, and serving them web pages, either static or dynamic
pages.

\item A computer program that provides the functionality described
in the first sense of the term.
\end{enumerate}

Web server also works based on several standards, such as [2]:
HTTP response to HTTP Request, Logging, Configurability,
Authentication, Handling Static and Dynamic Contents, Modular
Support, Virtual Hosts

\subsection{Erlang/OTP}

Erlang is a concurrent programming language with a functional
core. By this we mean that the most important property of the
language is that it is concurrent and that secondly, the
sequential part of the language is a functional programming
language. Concurrent means that the language has focus on how to
makes multiple executions threads to run and do computational work
together. In Erlang, these execution threads are called processes.
The sequential sub-set of the language expresses what happens from
the point it time where a process receives a message to the point
in time when it emits a message.

The early version of Erlang was developed by Ericsson Computer
Science Laboratory in 1985. During that time, Ericsson couldn't
find an appropriate language that has high performance in
concurrency especially for telecommunication applications
programming (for switching, trunking, etc), so they developed
their own language. OTP stands for Open Telecom Platform, OTP was
developed by Ericsson Telecom AB for programming next generation
switches and many Ericsson products are based on OTP. OTP includes
the entire Erlang development system together with a set of
libraries written in Erlang and other languages. OTP was
originally designed for writing telecoms application but has
proved equally useful for a wide range of non-telecom that have
concurrent, distributed, and also fault tolerant applications. In
1998 Ericsson released Erlang and the OTP libraries as open
source. Now, Erlang/OTP has reached the R11 version.

\section{Why We Use Erlang ?}

The simple answer to the question above that is we need
concurrency in the benchmark application. The application must be
able to generate multiple users to do some stress tests to the web
server.

But, there are some good features in Erlang, and even the other
languages don't have these features. Some of these Erlang features
are described below:

\begin{enumerate}
\item In Erlang processes are light weight.

\item Not only are Erlang processes light-weight, but also we can
create many hundreds of thousands of such processes without
noticeably degrading the performance of the system (unless of
course they are all doing something at the same time)[5].

\item In Erlang, processes share no data and the only way in which
they can exchange data is by explicit message passing.
``dangling'' pointers are very difficult to program in the
presence of hardware failures - we took the easy way out, by
disallowing all such data structures. [4]

\item In Erlang, processes scheduling operation is done by its own
virtual machine, so Erlang didn't inherit the underlying operating
system processes scheduling.

\item Real time. Erlang is intended for programming soft real-time
systems where response times in the order of milliseconds are
required.[4]

\item  Continuous operation.

\item  Automatic Memory management.

\item  Distribution.
\end{enumerate}

\subsection{Concurrent and Distributed Erlang}

Concurrent in Erlang involves processes creation and deletion. In
order to create a new process in Erlang, we use BIF (Built In
Function) \textit{spawn/3} :
\texttt{spawn(modulename,functionname,argumentlists)}

or

\noindent
\texttt{pid\_variabe=spawn(modulename,functionname,\\
argumentlists)}

\begin{figure}[htbp]
\includegraphics[width=52.0mm, height=37.7mm]{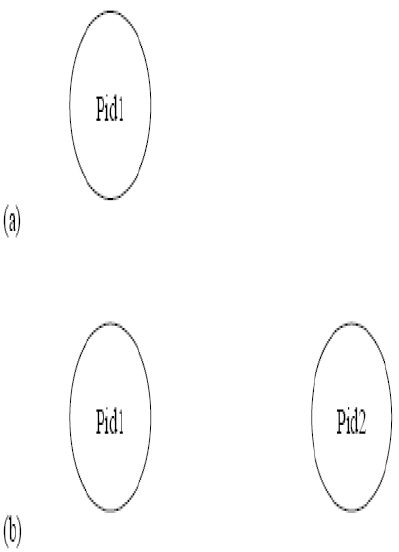}
\caption{Process Creation Illustration}
\label{image1}
\end{figure}

The illustration of process creation can be seen in figure
\ref{image1}.

A process which no longer need by the system will be automatically
shutdown/delete by the virtual machine (Erlang Runtime
System/ERTS). Meanwhile, the message parsing mechanism can be done
by these codes :

\begin{tabular}{|p{1.2in}|} \hline
Pid ! Message, \\ \hline
..... \\ \hline
receive \\ \hline
Message1 -$>$ \\ \hline
Actions1; \\ \hline
Message2 -$>$ \\ \hline
Actions2; \\ \hline
... \\ \hline
after Time -$>$ \\ \hline
TimeOutActions \\ \hline
end \\ \hline
\end{tabular}

\begin{figure}[htbp]
\includegraphics[width=55.7mm, height=34.1mm]{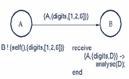}
\caption{Message Parsing Illustration}
\label{image2}
\end{figure}

The illustration for message parsing can be seen in figure
\ref{image2}.

Besides the example that has been shown above, in Erlang we can
also use Behaviour in OTP Design Principles to create, delete and
do message parsing between processes. A distributed Erlang system
is a number of Erlang Runtime System (we called them nodes) that
communicated each other by using message parsing with pid (process
indentifier) through TCP/IP sockets transparently. A node must be
given a name before it can communicate each other. The name is
either a long name or a short name like the examples below:
\begin{verbatim}
\$erl -name dilbert[long name]
(dilbert@uab.ericsson.se)1$>$
@erl-sname dilbert[short name]
(dilbert@uab)1$>$
\end{verbatim}

A simple distribution in Erlang can be done by these codes:

\begin{tabular}{|p{1.2in}|} \hline
...  \\ \hline
Pid = spawn(Fun@Node)  \\ \hline
...  \\ \hline
alive(Node)  \\ \hline
...  \\ \hline
not\_alive(Node)  \\ \hline
 \\ \hline
\end{tabular}

\subsection{Benchmark}

The term benchmark can be described as:

\begin{enumerate}
\item A group of parameters in which products (software or
hardware) can be measured the performance according to these
parameters.

\item A computer program designed to measure the performance of
software or hardware according to certain parameters.

\item A group of performance criteria that must be complied by
software or hardware.
\end{enumerate}

Web server benchmark means that a benchmark activity is made to
the web server to measure its performance based on several
parameters and using certain computer program to do this activity.

\section{Design}

\subsection{WiiBench Concepts}

WiiBench is divided into two main parts, the controller section
and the launcher section.

Controller is the core of the application. It has several duties,
such as read the benchmark scenario file, configure the program
based on the scenario, initialize the launcher section, gather the
benchmark results from local and remote Erlang node where the
launcher runs and write them in a log file (later the log file
will be used to generate a report page for the sysadmin).
Controller also has function as a timer which act as timing for
user inters arrival to the server.

Launcher generates a number of user based on the scenario,
initialize them and start the benchmark by sending requests to the
web server. The clients also gather the benchmark result and send
them to the controller.

\begin{figure}[htbp]
\includegraphics[width=70mm, height=79.0mm]{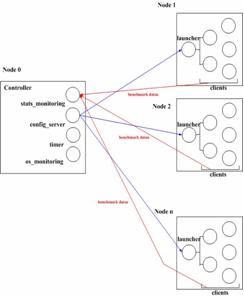}
\caption{Illustration for WiiBench Concepts} \label{image3}
\end{figure}

The illustration for WiiBench Concepts can be seen in figure
\ref{image3}

Several parameters that can be measured by this application are:

\begin{enumerate}
\item Number of Requests per second.

\item Simultaneous users that can be served by the server (per
second).

\item Time that needs to connect for the client so it can be
connected to server.

\item Number of user that can be served during the duration of
benchmark.

\item Time that needs for a user so it can receive a full page of
document/web page according to the request.

\item Time that needs to complete a session (a group of requests)
as described in the scenario.

\item Network throughput.

\item  HTTP Status (200, 404).
\end{enumerate}

The application also has a report generator that written in PERL
and using GNUPLOT to generate graphs based on the benchmark
results from the log file.

\subsection{The Scenario Files}

The benchmark scenario file is written using XML and consists of several sections:

\begin{enumerate}
\item The server section, where the user (sysadmin) describe the
IP address of the server that he/she wants to benchmark.

\item The client section, in this section, user can write the IP
addresses where the launcher section starts and generate a numbers
of clients.

\item Inters arrival phase and benchmark duration section.

\item The simulated user agents (web browser) section.

\item The session and request section.
\end{enumerate}

Each of the section describes above can be modified according to user necessity.

\section{Implementation}

\subsection{Hardware and Software Specifications}

WiiBench is implemented by using a simple topology consists of two
computers and a server across Local Area Network in Kapuk Valley,
Margonda, Depok. The two computers using an Intel Celeron
Processor (1.8 GHz and 2.28 GHz) and running Linux operating
System (SuSE 10.0 and Slackware Linux 11.0), Open SSH v2,
Erlang/OTP R11, PERL v5.8, and also BASH Shell. Both of them also
connected to the TCP/IP network and within Kapuk-Valley domain
(hostname mobile and posen-asik). The server using an Intel
Pentium II 400 MHz processor with 768 megabytes SDRAM memories and
running Slackware Linux 11.0 with Apache 2.0.55 web server.

\subsection{Testing and Implementation Process}

In this implementation process, we're going to make a benchmark scenario with these parameters:

\begin{enumerate}
\item Total clients to generate: 600 clients.

\item Benchmark duration: 10 minutes.

\item Client inters arrival phase: 1 second.

\item 300 clients will be generating in host mobile, meanwhile the
other 300 clients will be generate in host posen-asik.
\end{enumerate}

Before we run the program, we must generate a pair of
authentication key (public and private) for passwordless
authentication using SHH in order to get the Erlang nodes to
communicate each other.

\texttt{[root@mobile\~{}]\#ssh-keygen -t rsa}

\texttt{Enter file in which to save the key (/root/.ssh/id\_rsa):}

\texttt{Enter passphrase (empty for no passphrase):}

\texttt{Enter same passphrase again:}

\texttt{Your identification has been saved in /root/.ssh/id\_rsa.}

\texttt{Your public key has been saved in /root/.ssh/id\_rsa.pub.}

\texttt{The key fingerprint is:}

\texttt{ec:30:2c:c9:e0:0a:92:48:3c:e5:5a:f3:7c:69:d8:92
root@mobile.myownlinux.org}

\texttt{[root@mobile\~{}]\#scp /root/.ssh/id\_rsa.pub
root@posen-asik:/root/.ssh/}

\texttt{[root@posen-asik]\#echo .ssh/id\_rsa.pub $>$$>$
.ssh/authorized\_keys}

After the process above, we can start the benchmark process by
executing the shell script to initialize and start the
application.

\texttt{[root@mobile\~{}]\#/usr/local/bin/wii start}

Several results from the important parameters to examine by the sysadmin are listed in the table below.

\textbf{Table 4.1 Benchmark Results}

\begin{tabular}{|p{0.9in}|p{1.0in}|} \hline
\textbf{Parameters} & \textbf{Result} \\ \hline
Request per Second & 3.4 requests per second \\ \hline
Connection per Second & 1.8 connections per second \\ \hline
Page loaded per second & 1.8 pages per second \\ \hline
Total user served & 595 of 600 users \\ \hline
\end{tabular}

\textbf{}

\section{Conclusions}

After our studies we show that in Erlang:

\begin{enumerate}
\item Processes are light weight. Not only are Erlang processes
light-weight, but also we can create many hundreds of thousands of
such processes without noticeably degrading the performance of the
system (unless of course they are all doing something at the same
time),

\item  Processes share no data and the only way in which they can
exchange data is by explicit message passing. Erlang message never
contain pointers to data and since there is no concept of shared
data, each process must work on a copy of the data that it needs.
All synchronization is performed by exchanging messages.

\item Processes scheduling operation is done by its own virtual
machine.

\item Processes are real time. Erlang is intended for programming
soft real-time systems where response times in the order of
milliseconds are required.

\item Processes code has continuous operation. Erlang has
primitives which allow code to be replaced in a running system and
allow old and new versions of code to execute at the same time.

\item Processes operation use automatic memory management. Memory
is allocated automatically when required, and deallocated when no
longer used.

\item All interaction between processes is by asynchronous message
passing. Distributed systems can easily be built.
\end{enumerate}

By examine the results based on the important parameters; we hope
that the syadmin and netadmin can make fine tuning to their
server.

\section*{Acknowledgment}
The authors would like to thank Gunadarma University's Research
Council

\begin{biography}{Achmad B. Mutiara}
(M'05) was born in Jakrata, in 1967. He received the B.Sc.\ degree
in Informatics Engineering from Gunadarma University, Depok, in
1991, and the M.S.\ degree and Ph.D\ degree in computational
material physics from Goettingen University, Germany, in 1997 and
2000.
\end{biography}

\begin{biography}{T.A. Sabastian}
He received the B.Sc.\ degree in Informatics Engineering from
Gunadarma University, Depok, in 2007.
\end{biography}

\end{document}